\documentclass[letterpaper]{article} 
\usepackage{aaai25}  
\usepackage{times}  
\usepackage{helvet}  
\usepackage{courier}  
\usepackage[hyphens]{url}  
\usepackage{graphicx} 
\usepackage{natbib}  
\usepackage{caption} 
\usepackage{algorithm}
\usepackage{algorithmic}

\usepackage{newfloat}
\usepackage{listings}
\DeclareCaptionStyle{ruled}{labelfont=normalfont,labelsep=colon,strut=off} 
\floatstyle{ruled}
\newfloat{listing}{tb}{lst}{}
\floatname{listing}{Listing}
\usepackage{colortbl}
\definecolor{cvprblue}{rgb}{0.21,0.49,0.74}
\definecolor{Gray}{gray}{0.85}
\newcolumntype{a}{>{\columncolor{Gray}}c}

\newcommand{\ie}{i.e.}

\usepackage{multirow} 
\usepackage{amsmath}
\usepackage{graphicx}
\usepackage{booktabs} 
\usepackage{amssymb}
\usepackage{sidecap, caption}
\usepackage{dsfont}
\usepackage{enumitem}

\newcommand\best[1]{\textbf{#1}}

\title{Protecting Model Adaptation from Trojans in the Unlabeled Data}
\author{
    Lijun Sheng\textsuperscript{\rm 1,2}, Jian Liang\textsuperscript{\rm 2,3}\thanks{To whom correspondence should be addressed.}, Ran He\textsuperscript{\rm 2,3}, Zilei Wang\textsuperscript{\rm 1}, Tieniu Tan\textsuperscript{\rm 4,2,3}
}
\affiliations{
    \textsuperscript{\rm 1} University of Science and Technology of China \\
    \textsuperscript{\rm 2} NLPR \& MAIS, Institute of Automation, Chinese Academy of Sciences \\
    \textsuperscript{\rm 3} University of Chinese Academy of Sciences
    \textsuperscript{\rm 4} Nanjing University \\
    {slj0728@mail.ustc.edu.cn, liangjian92@gmail.com, zlwang@ustc.edu.cn, \{rhe, tnt\}@nlpr.ia.ac.cn} \\
}

\begin{document}

\maketitle

\begin{abstract}
Model adaptation tackles the distribution shift problem with a pre-trained model instead of raw data, which has become a popular paradigm due to its great privacy protection.
Existing methods always assume adapting to a clean target domain, overlooking the security risks of unlabeled samples.
This paper for the first time explores the potential trojan attacks on model adaptation launched by well-designed poisoning target data.
Concretely, we provide two trigger patterns with two poisoning strategies for different prior knowledge owned by attackers.
These attacks achieve a high success rate while maintaining the normal performance on clean samples in the test stage.
To defend against such backdoor injection, we propose a plug-and-play method named {\sc DiffAdapt}, which can be seamlessly integrated with existing adaptation algorithms.
Experiments across commonly used benchmarks and adaptation methods demonstrate the effectiveness of {\sc DiffAdapt}.
We hope this work will shed light on the safety of transfer learning with unlabeled data.
\end{abstract}

\begin{links}
\link{Code}{https://github.com/TomSheng21/DiffAdapt}
\end{links}

\section{Introduction}

Over recent years, deep neural networks \citep{krizhevsky2012imagenet, he2016deep, dosovitskiy2020image} have gained substantial research interest and demonstrated remarkable capabilities across various tasks.
However, distribution shift \citep{saenko2010adapting} between the training set and deployment environment inevitably arises, leading to a significant drop in performance.
To solve this issue, researchers propose domain adaptation \citep{ben2010theory, ganin2016domain, long2018conditional} to improve the performance on unlabeled target domains by utilizing labeled source data.
As privacy awareness grows, source providers restrict user's access to raw data.
Instead, model adaptation \citep{liang2020we}, a novel paradigm only accessing pre-trained source models, has gained popularity \citep{liang2020we, li2020model}.
Since its proposal, model adaptation has been extensively investigated across various visual tasks, including semantic segmentation \citep{fleuret2021uncertainty, liu2021source} and object detection \citep{li2021free, huang2021model}.

\begin{figure}[t]
    \centering
    \includegraphics[width=0.47\textwidth]{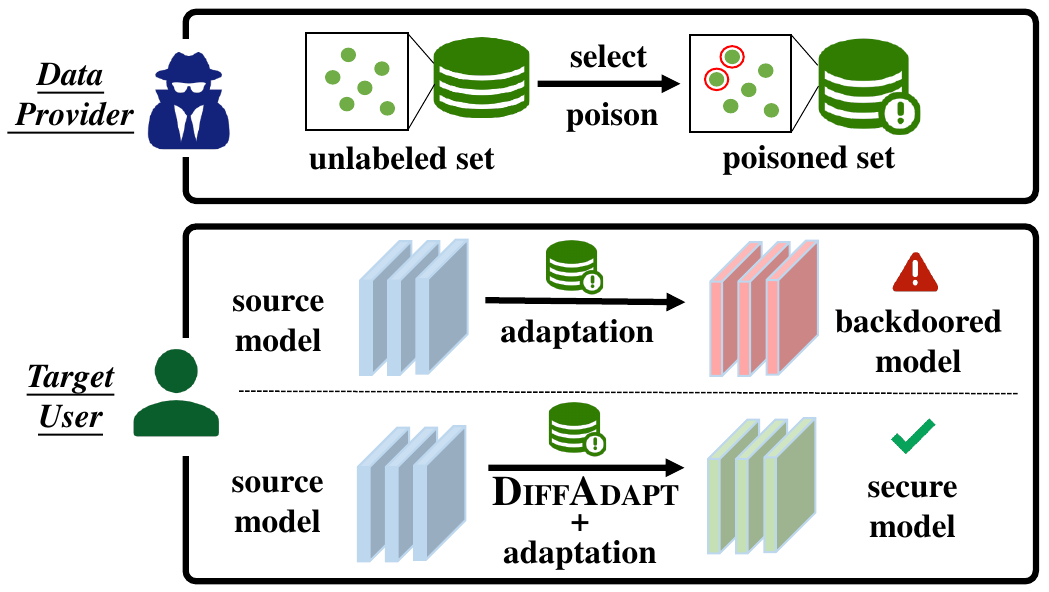}
    \caption{Backdoor attack and defense on model adaptation. With well-poisoned unlabeled data from malicious providers, target users suffer from the risks of backdoor injection. We propose {\sc DiffAdapt}, a defense method against backdoor injection without sacrificing clean performance.}
  \label{fig:intro}
\end{figure}

Security problems in model adaptation are always ignored, only two recent works \citep{sheng2023adaptguard, ahmed2023ssda} reveal its vulnerability to the neural trojans (also known as backdoors) \citep{gu2017badnets, chen2017targeted} embedded in the source model.
A distillation framework \citep{sheng2023adaptguard} and a model compression scheme \citep{ahmed2023ssda} are proposed to eliminate threats from suspicious source providers, respectively.
In this paper, we raise a similar question: Can we trust the unlabeled target data?
Different from the source model, injecting trojans through unlabeled data faces several significant challenges.
It is difficult for unsupervised algorithms to directly establish a strong connection between the trigger and the target class through poisoning unlabeled data.
Nevertheless, we find that well-poisoned unlabeled datasets still achieve successful backdoor attacks on adaptation algorithms, as illustrated in Fig.~\ref{fig:intro}. 

We decompose unsupervised backdoor injection into two parts: trigger design and poisoning strategy.
First, a non-optimization-based trigger and an optimization-based trigger are introduced.
We adopt the Hello Kitty trigger in Blended \citep{chen2017targeted} as the non-optimization-based trigger.
The optimization-based one is an adversarial perturbation \citep{poursaeed2018generative} calculated with a surrogate model.
As for the poisoning sample selection, we provide two strategies for different prior knowledge owned by attackers.
In cases where the attackers have ground truth labels, samples belonging to the target class are directly selected.
When labels cannot be accessed, samples to be poisoned are selected by querying the open-source CLIP model \citep{radford2021learning}.
Please note that after the poisoning stage, attackers release the unlabeled version of the dataset for downstream adaptation users.
Experimental results have shown that the collaboration of designed triggers and poisoning strategies achieves successful backdoor attacks.

To defend model adaptation against the backdoor threat, we propose a plug-and-play method called {\sc DiffAdapt}.
{\sc DiffAdapt} eliminates the mapping between the backdoor trigger and the target class by neglecting potentially poisoning target samples during optimization.
First, we train a potential risk target model with unlabeled data following the common model adaptation algorithms (e.g., SHOT \citep{liang2020we}, NRC \citep{yang2021exploiting}).
Since poisoning samples have both semantic features and backdoor triggers connected with the target class, they tend to be less sensitive to random noise perturbation.
We calculate the distance between the output prediction of the original and perturbed version for every unlabeled data to form the sample weight.
Finally, the sample weight is averaged by the distance of all samples with the same pseudo label and a new secure target model is trained with the sample weight to avoid being injected with backdoors during unsupervised adaptation.
Since no requirements for loss functions and network architectures, {\sc DiffAdapt} can seamlessly integrate with existing adaptation algorithms.
In the experiment section, we demonstrate the effectiveness of {\sc DiffAdapt} on two popular model adaptation methods (\ie, SHOT \citep{liang2020we}, and NRC \citep{yang2021exploiting}) across three frequently used datasets (\ie, Office \citep{saenko2010adapting}, OfficeHome \citep{venkateswara2017deep}, and DomainNet \citep{peng2019moment}).
Our contributions are summarized as follows:
\begin{itemize}[leftmargin=20pt]
\item We explore backdoor attacks on model adaptation through poisoning unlabeled target data. To the best of our knowledge, this is the first attempt at unsupervised backdoor attacks during adaptation tasks.
\item We provide two poisoning strategies coupled with two trigger patterns capable of successfully embedding neural trojans into existing adaptation algorithms.
\item We propose {\sc DiffAdapt}, a flexible plug-and-play defense method against potential backdoor attacks while maintaining task performance on clean data.
\item Extensive experiments involving two model adaptation methods across three benchmarks demonstrate the effectiveness of {\sc DiffAdapt}.
\end{itemize}

\section{Related Work}
\subsection{Model Adaptation}
Model adaptation \citep{liang2020we, yang2021exploiting, ding2023proxymix, liang2021source}, aims to transfer knowledge from a pre-trained source model to an unlabeled target domain, which is also called source-free domain adaptation or test-time domain adaptation \citep{liang2024comprehensive, yu2023benchmarking}.
SHOT \citep{liang2020we} first exploits this paradigm and employs information maximization loss and self-supervised pseudo-labeling to achieve source hypothesis transfer.
NRC \citep{yang2021exploiting} captures target feature structure and promotes label consistency among high-affinity neighbor samples.
Some methods \citep{li2020model, zhang2022divide, tian2021vdm} attempt to estimate the source domain or select source-similar samples to benefit knowledge transfer.
Existing works also discuss many variants of model adaptation, such as black-box adaptation \citep{liang2022dine, zhang2023black}, open-partial \citep{liang2021umad}, and online \citep{wang2020tent, yu2024stamp} scenarios.

With widespread attention on the security topic, a series of works \citep{agarwal2022unsupervised, li2021imbalanced, sheng2023adaptguard, ahmed2023ssda} have studied the security of model adaptation.
A robust adaptation method \citep{agarwal2022unsupervised} is proposed to improve the adversarial robustness of model adaptation.
ISFDA \citep{li2021imbalanced} focuses on adaptation on class-imbalanced target dataset.
AdaptGuard \citep{sheng2023adaptguard} investigates the vulnerability to image-agnostic attacks launched by the source side and introduces a model processing defense framework.
SSDA \citep{ahmed2023ssda} proposes a model compression scheme against source backdoor attacks.
However, this paper focuses on the trojan attack on model adaptation through unlabeled poisoning data, which has not been studied so far.

\subsection{Backdoor Attack and Defense}
Backdoor attack \citep{gu2017badnets, wu2022backdoorbench, li2022backdoor, zhang2023red, liao2024imperceptible} is an emerging security topic to plant a neural trojan or a backdoor associated with a trigger pattern in deep neural networks.
Many well-designed backdoor triggers are proposed to achieve trojan injection.
BadNets \citep{gu2017badnets} utilizes a pattern of bright pixels to attack digit classifiers and street sign detectors.
Blended \citep{chen2017targeted} achieves a strong invisible backdoor attack by mixing samples with a cartoon image.
ISSBA \citep{li2021invisible} proposes a sample-specific trigger generated through an encoder-decoder network.
In addition to poisoning-based solutions, some methods enhance their attack effects by controlling the training process \citep{nguyen2021wanet, doan2021lira}.

Recently, backdoor attacks have been studied in diverse scenarios besides supervised learning.
Some works \citep{saha2022backdoor, li2023embarrassingly} explore the backdoor attacks for victims who deploy self-supervised methods on unlabeled datasets.
A repeat dot matrix trigger \citep{shejwalkar2023perils} is designed to attack semi-supervised learning methods by poisoning unlabeled data.
Backdoor injection \citep{chou2023backdoor, chou2024villandiffusion} also works on diffusion models \citep{dhariwal2021diffusion}.
Our work tries to launch the backdoor attack on model adaptation via poisoning unlabeled data, evaluating the danger of backdoor attacks from a new perspective.

With the emergence of trojan attack methods, various backdoor defense methods \citep{liu2018fine, wang2019neural, guan2022few, guan2024backdoor} are proposed alternately.
Fine-Pruning \citep{liu2018fine} finds that a combination of pruning and fine-tuning can effectively weaken backdoors.
NAD \citep{li2021neural} optimizes the backdoored model using a distillation loss with a fine-tuned teacher model.
ANP \citep{wu2021adversarial} identifies and prunes backdoor neurons that are more sensitive to adversarial neuron perturbation.
CLP \citep{zheng2022data} removes risky channels with a high channel Lipschitz constant in a data-free way.
However, those defense methods are deployed on in-distribution data and most of them require labeled samples, which are impractical for model adaptation.

\section{Backdoor Attack on Model Adaptation}

In this section, we focus on the backdoor attack on model adaptation through unsupervised poisoning.
First, we review the model adaptation framework and introduce the challenge and attacker's knowledge of injecting trojans during adaptation. 
Subsequently, we decompose backdoor embedding into trigger design and data poisoning strategy, providing a detailed discussion respectively.

\subsection{Preliminary Knowledge}
\label{sec:pre}

\textbf{Model adaptation} \citep{liang2020we}, also known as source-free domain adaptation, aims to adapt a pre-trained source model $f_s$ to a related target domain.
Two domains share the same label space but follow different distributions with a domain gap.
Model adaptation methods employ unsupervised learning techniques with the source model $f_s$ and unlabeled data $\mathcal{D}_t = \{x_i^t\}_{i=1}^{N_t}$ to obtain a model $f_t$ with better performance on the target domain.

\textbf{Challenges of backdoor attacks on model adaptation.}
Unlike conventional backdoor embedding methods, backdoor attacks on model adaptation encounter several additional challenges.

Previous attackers always achieve backdoor embedding on supervised learning by adding the trigger on some samples and modifying their labels with the target class.
Supervised victim learners using the poisoned dataset will capture the mapping from the trigger to the target class.
However, for model adaptation algorithms, attackers are restricted to poison unlabeled data which establishes a weak connection.
Moreover, the weak optimization ability of unsupervised fine-tuning also makes implanting new features very challenging.

\textbf{The attacker's knowledge.}
In the scenario of backdoor attacks on model adaptation, attackers are allowed to control only target data, and in extremely challenging cases, only the data supply of the target class.
As the target data owner, the attacker may have ground truth labels or obtain pseudo labels through the open-source basic model (e.g., CLIP \citep{radford2021learning}).
Last but not least, the attacker is not allowed to access the source model and have no knowledge about the downstream adaptation learners.

\subsection{Backdoor Triggers}
\label{sec:trigger}
To make adaptation algorithms capture the mapping from the trigger to the target class, we utilize the triggers with semantic information.
Triggers with semantic information will be extracted by the pre-trained source model with a higher probability.
Based on this requirement, we introduce a non-optimization-based trigger and an optimization-based perturbation trigger in the following.

\noindent $\rhd$ \textbf{A non-optimization-based (Blended) trigger.}
Blended \citep{chen2017targeted} is a strong backdoor attack technique that blends the samples with a Hello Kitty image.
Blended trigger satisfies the requirements and no additional knowledge is required, making it a suitable choice as our non-optimization-based trigger.

\noindent $\rhd$ \textbf{An optimization-based (perturbation) trigger.}
In addition to the hand-crafted trigger, we introduce an optimization-based method for trigger generation.
Initially, a surrogate model is trained on the target dataset $\mathcal{{D}}_t = \{x_i^t, {{y}}_i\}_{i=1}^{N_t}$ using a cross-entropy loss function.
With the surrogate model and the target data, we compute the universal adversarial perturbations \citep{poursaeed2018generative} for the target class which leads to the misclassification of the majority of samples.
The perturbation has misleading semantics and is the same size as input samples which makes it a great optimization-based trigger.
It is worth noting that the architecture of the surrogate model and the source model are always different, and the perturbation will not achieve such a high attack success rate on the source model due to the weak transferability between different architectures.

\subsection{Data Poisoning}
\label{sec:poison}
Previous backdoor attack methods employ data poison with a random sampling strategy.
Due to the unsupervised nature of adaptation algorithms, attackers are unable to establish a connection between the trigger and the target class explicitly.
Hence, a well-designed poison set selection strategy becomes critical backdoor embedding.
Attackers are allowed to access either ground truth or open-source basic model predictions for the poisoning data selection.
We provide a selection strategy for each condition below.

\noindent $\rhd$ \textbf{Ground-truth-based selection strategy.}
When attackers hold the ground truth labels $\{y_i^t\}_{i=1}^{N_t}$ of all samples, samples belonging to the target class $y_t$ are simply selected to construct a poisoning set $\mathcal{D}_t^{poison} = \{x_i^t\}_{i=1}^{P}$.
To avoid interference for backdoor embedding, samples in other classes remain unchanged.

\noindent $\rhd$ \textbf{Pseudo-label-based selection strategy.}
With no access to the ground truth labels, attackers first obtain pseudo-labels for all target data from the open-source basic model CLIP \citep{radford2021learning}.
Then, a base poisoning set $\mathcal{D}_t^{pl} = \{x_i^t\}_{i=1}^{P}$ consists of samples belonging to the target class $y_t$.
In order to strengthen the poisoning set, attackers can choose to continually select samples outside of $\mathcal{D}_t^{pl}$ but with a high prediction probability for the target class $y_t$, creating a supplementary set $\mathcal{D}_t^{supp} = \{x_i^t\}_{i=1}^{P'}$.
The final poisoning set $\mathcal{D}_t^{poison}$ is the union of the above two sets: $\mathcal{D}_t^{poison} = \mathcal{D}_t^{pl} \cup \mathcal{D}_t^{supp}$.

\begin{figure*}[t]
    \centering
    \includegraphics[width=0.8\linewidth]{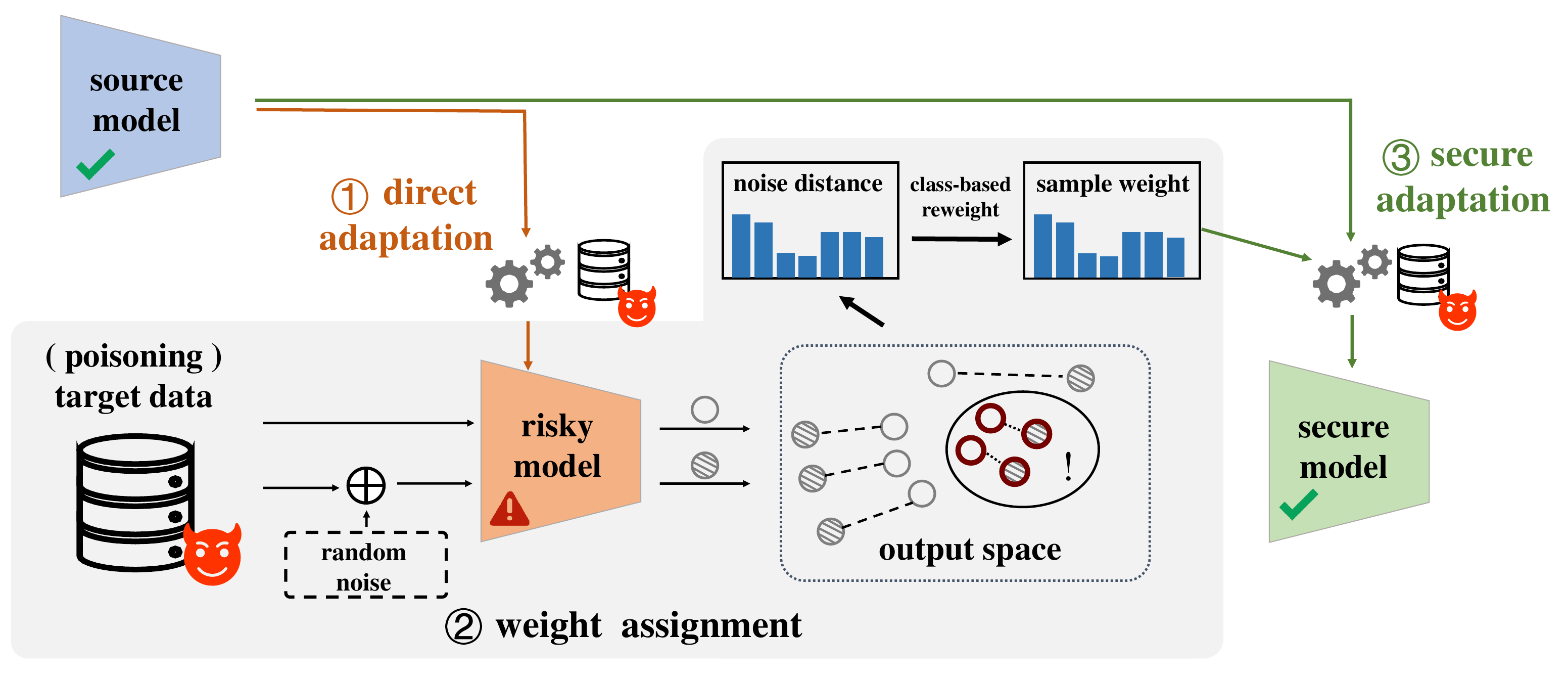}
    \caption{The framework of defense method {\sc DiffAdapt}. We train a potentially risky target model and obtain the distance between the output prediction of the original and perturbed version for every unlabeled data. The sample weight is averaged by the distance of all samples with the same pseudo label and a new secure target model is trained with the sample weight.}
    \label{fig:defense}
\end{figure*}

\section{{\sc DiffAdapt}: A Secure Adaptation Method Against Backdoor Attacks}

From the previous section, we learn an incredible fact: malicious target data providers can achieve a backdoor embedding on model adaptation algorithms through unsupervised poisoning.
We introduce a straightforward defense method named {\sc DiffAdapt} to mitigate such a risk.
{\sc DiffAdapt} is designed to defend against potential backdoor attacks while preserving the adaptation performance in the target domain.
The framework of {\sc DiffAdapt} is illustrated in Fig.~\ref{fig:defense}.

The main idea inside {\sc DiffAdapt} is intuitive, assigning low weights to samples that may contain backdoor triggers, instead of the uniform weights in existing adaptation methods.
The key is to obtain accurate weights, that is, identify samples containing backdoor triggers.
For risky target models that may contain backdoors, the outputs of poisoned samples in the unlabeled training set will be determined by their category semantics and the trigger.
Since these samples have both features connected with the target class, they tend to be less sensitive to random noise perturbation, that is, there will be little change between the output of the original image and the perturbed version.
Therefore, {\sc DiffAdapt} assigns weights to samples based on their sensitivity to noise on potentially risky target models.

Here, we provide a detailed outline of the procedures involved in {\sc DiffAdapt}.
Firstly, a potentially risky target model $f_t^{risk}$ is trained using the existing model adaptation algorithm with unlabeled target dataset $\mathcal{D}_t$.
Note that ``potentially'' means that the target model $f_t^{risk}$ may be secure.
Our defense method does not assume that the unlabeled target domain has to contain poisoned samples.
To obtain the sensitivity of each target sample, we randomly generate a Gaussian noise $\epsilon$ and construct a perturbed target domain dataset.
The distance between the predictions of the original image and the perturbed version is calculated as follows:
\begin{equation}
    \delta(x_{t,i}) = \lVert {f_t^{risk} (\widetilde{x}_{t,i} )} - {f_t^{risk} ( x_{t,i} )} \rVert ,
    \label{eq:distance}
\end{equation}
where $\widetilde{x}_{t,i} = x_{t,i} + \epsilon$ represents the perturbed image.
Since the risky target model $f_t^{risk}$ has incorporated the knowledge of unlabeled training data, a certain proportion of the data tends to be less sensitive to noise like the poisoned samples.
In addition, categories that are not related to backdoors contain some highly sensitive outlier samples, and the model's high dependence on them will decrease the stability of the adaptation process.
Therefore, to focus on low-weight clean samples while reducing the dependence on high-weight samples, {\sc DiffAdapt} reassigns sample weights according to pseudo labels.
The final weight $w_{i}$ is the average among all samples with the same pseudo label:
\begin{equation}
    w(x_{t,i}) = \frac{\sum_{j=1}^{N_t}{\mathds{1}(\hat{y}_j=\hat{y}_i)}\ \delta(x_{t,j})}{\sum_{j=1}^{N_t}{\mathds{1}(\hat{y}_j=\hat{y}_i)}},
    \label{eq:weight}
\end{equation}
where $ \hat{y}_j = \textbf{argmax}_c f_t^{risk}(x_{t,j})_c$ represents the pseudo label provided by the risk target model.
Finally, the secure target domain model is obtained by retraining using the adaptation algorithm.
The objective corresponding to each sample in the adaptation process will be replaced by a weighted version with $w_{i}$:
\begin{equation}
    L (x_{t,i}) = \mathbb{E}_{x_{t,i} \in \mathcal{D}_t} {w(x_{t,i}) \ l (x_{t,i})},
    \label{eq:reweight loss}
\end{equation}
where $l (x_{t,i})$ refers to the loss calculated on the sample $x_{t,i}$, for example, self-training loss and entropy minimization loss in SHOT \citep{liang2020we} and class-consistency loss in NRC \citep{yang2021exploiting}.

\textbf{Discussion.}
Since {\sc DiffAdapt} has no requirements on model adaptation algorithms and network architectures, it can be used as a plug-and-play defense strategy simply combined with existing adaptation algorithms (e.g., SHOT \citep{liang2020we}, and NRC \citep{yang2021exploiting}).
Besides effectively defending against test-time backdoor attacks, {\sc DiffAdapt} maintains the adaptation performance on the clean data.

\setlength{\tabcolsep}{4.0pt}
    \begin{table*}[!t]
        \centering
        \resizebox{0.95\textwidth}{!}{
            \begin{tabular}{l|rr|rr|rr|rr|rr|rr|rr|rr|rr|rr}
            \toprule
            & \multicolumn{10}{c|}{\textbf{SHOT} \citep{liang2020we}} & \multicolumn{10}{c}{\textbf{NRC} \citep{yang2021exploiting}} \\
            \midrule
            \multirow{2}{3em}{Task} & \multicolumn{2}{c|}{A$\to$W} & \multicolumn{2}{c|}{D$\to$A} & \multicolumn{2}{c|}{D$\to$W} & \multicolumn{2}{c|}{W$\to$A} & \multicolumn{2}{c|}{Avg} & \multicolumn{2}{c|}{A$\to$W} & \multicolumn{2}{c|}{D$\to$A} & \multicolumn{2}{c|}{D$\to$W} & \multicolumn{2}{c|}{W$\to$A} & \multicolumn{2}{c}{Avg} \\ 
            & ACC & ASR & ACC & ASR & ACC & ASR & ACC & ASR & ACC & ASR & ACC & ASR & ACC & ASR & ACC & ASR & ACC & ASR & ACC & ASR  \\
            \midrule
            Source Only & 77.4 & \multicolumn{1}{c|}{-} & 62.0 & \multicolumn{1}{c|}{-} & 95.0 &  \multicolumn{1}{c|}{-}  & 63.6 &  \multicolumn{1}{c|}{-}  & 74.5 & \multicolumn{1}{c|}{-}  & 77.4 & \multicolumn{1}{c|}{-} & 62.0 & \multicolumn{1}{c|}{-} & 95.0 & \multicolumn{1}{c|}{-}  & 63.6 & \multicolumn{1}{c|}{-} & 74.5 & \multicolumn{1}{c}{-}  \\
            No Poisoning & 92.5 & \multicolumn{1}{c|}{-} & 76.4 & \multicolumn{1}{c|}{-} & 97.5 & \multicolumn{1}{c|}{-} & 76.4 &  \multicolumn{1}{c|}{-}  & 85.7 & \multicolumn{1}{c|}{-}  & 93.7 & \multicolumn{1}{c|}{-} & 78.3 & \multicolumn{1}{c|}{-}  & 98.7 & \multicolumn{1}{c|}{-}   & 77.1 & \multicolumn{1}{c|}{-} & 87.0 & \multicolumn{1}{c}{-} \\
            \midrule
            \midrule
            Poisoning & 91.2 & 41.9 & 75.8 & 48.4 & 98.1 & 60.0 & 76.4 & 40.9 & 85.4 & 47.8 & 92.5 & 58.1 & 76.6 & 80.3 & 98.1 & 90.3 & 77.4 & 82.3 & 86.1 & 77.7 \\
            + {CLP} & 90.6 & 28.4 & 76.0 & \best{27.3} & 97.5 & 52.9 & 75.0 & \best{25.8} & 84.8 & 33.6 & 91.2 & 23.2 & 75.3 & 55.3 & 96.9 & 76.8 & 74.6 & 72.9 & 84.5 & 57.0 \\
            + {FP} & 88.1 & 39.4 & 75.3 & 51.2 & 93.1 & 48.4 & 73.4 & 43.1 & 82.5 & 45.5 & 86.8 & 63.2 & 73.7 & 78.3 & 90.6 & 78.1 & 72.7 & 82.1 & 80.9 & 75.4 \\
            \rowcolor{Gray}
            + {\sc DiffAdapt} & 87.4 & \best{9.0} & 74.4 & 47.0 & 98.7 & \best{34.8} & 72.7 & 29.7 & 83.3 & \best{30.1} & 85.5 & \best{0.0} & 75.1 & \best{49.5} & 98.1 & \best{53.6} & 72.3 & \best{23.8} & 82.8 & \best{31.7} \\
            \midrule
            \multicolumn{21}{c}{\textbf{Blended trigger $\upuparrows$} \ \ \ \ \ \ \ \ \ \textbf{Perturbation trigger $\downdownarrows$}}\\
            \midrule
            Poisoning & 91.2 & 64.5 & 76.6 & 88.2 & 97.5 & 87.7 & 74.8 & 86.0 & 85.0 & 81.6 & 92.5 & 38.7 & 76.4 & 59.9 & 98.1 & 76.8 & 78.2 & 55.1 & 86.3 & 57.6 \\
            + {CLP} & 91.2 & 27.7 & 75.1 & 61.1 & 96.9 & 38.1 & 75.0 & 41.1 & 84.5 & 42.0 & 91.2 & 6.5  & 75.1 & 25.8 & 97.5 & 22.6 & 74.6 & 5.5  & 84.6 & 15.1 \\
            + {FP} & 88.7 & 52.3 & 74.6 & 70.2 & 94.3 & 74.8 & 73.4 & 53.4 & 82.7 & 62.7 & 86.8 & 26.5 & 74.4 & 30.4 & 92.5 & 71.0 & 73.5 & 37.2 & 81.8 & 41.3 \\
            \rowcolor{Gray}
            + {\sc DiffAdapt} & 87.4 & \best{1.3} & 73.0 & \best{15.8} & 98.1 & \best{34.2} & 72.1 & \best{10.3} & 82.7 & \best{15.4} & 88.1 & \best{0.7} & 74.6 & \best{8.7} & 98.7 & \best{10.3} & 75.0 & \best{4.4} & 84.1 & \best{6.0} \\
            \bottomrule
            \end{tabular}
        }
        \caption{ACC (\%) and ASR (\%) of {\sc DiffAdapt} against backdoor attacks on \textbf{Office} \citep{saenko2010adapting} dataset for model adaptation (ResNet-50).}
    \label{tab:office} 
    \end{table*}

\setlength{\tabcolsep}{4.0pt}
    \begin{table*}[!t]
        \centering
        \resizebox{0.95\textwidth}{!}{
            \begin{tabular}{l|rr|rr|rr|rr|rr|rr|rr|rr|rr|rr}
            \toprule
            & \multicolumn{10}{c|}{\textbf{SHOT} \citep{liang2020we}} & \multicolumn{10}{c}{\textbf{NRC} \citep{yang2021exploiting}} \\
            \midrule
            \multirow{2}{3em}{Task} & \multicolumn{2}{c|}{A$\to$} & \multicolumn{2}{c|}{C$\to$} & \multicolumn{2}{c|}{P$\to$} & \multicolumn{2}{c|}{R$\to$} & \multicolumn{2}{c|}{Avg} & \multicolumn{2}{c|}{A$\to$} & \multicolumn{2}{c|}{C$\to$} & \multicolumn{2}{c|}{P$\to$} & \multicolumn{2}{c|}{R$\to$} & \multicolumn{2}{c}{Avg}  \\ 
            & ACC & ASR & ACC & ASR & ACC & ASR & ACC & ASR & ACC & ASR & ACC & ASR & ACC & ASR & ACC & ASR & ACC & ASR & ACC & ASR  \\
            \midrule
            Source Only & 60.2 &  \multicolumn{1}{c|}{-}  & 58.6 &  \multicolumn{1}{c|}{-} & 54.7 &  \multicolumn{1}{c|}{-} & 62.4 & \multicolumn{1}{c|}{-}  & 59.0 &  \multicolumn{1}{c|}{-}  & 60.2 & \multicolumn{1}{c|}{-}  & 58.6 & - & 54.7 & \multicolumn{1}{c|}{-} & 62.4 & \multicolumn{1}{c|}{-} & 59.0 & \multicolumn{1}{c}{-} \\
            No Poisoning & 71.3 & \multicolumn{1}{c|}{-} & 73.4 &  \multicolumn{1}{c|}{-}  & 67.2 &  \multicolumn{1}{c|}{-} & 69.5 &  \multicolumn{1}{c|}{-}  & 70.4 & \multicolumn{1}{c|}{-} & 71.1 &  \multicolumn{1}{c|}{-} & 72.9 & \multicolumn{1}{c|}{-}  & 64.6 & \multicolumn{1}{c|}{-} & 69.8 & \multicolumn{1}{c|}{-} & 69.6 & \multicolumn{1}{c}{-} \\
            \midrule
            \midrule
            Poisoning & 70.7 & 85.6 & 73.6 & 89.2 & 66.8 & 62.5 & 69.8 & 86.0 & 70.2 & 80.8 & 71.0 & 85.9 & 72.4 & 87.9 & 65.0 & 89.3 & 69.0 & 84.6 & 69.4 & 86.9 \\
            + {CLP} & 68.3 & 77.7 & 70.6 & 86.5 & 63.8 & 59.4 & 67.7 & 80.9 & 67.6 & 76.1 & 68.3 & 80.4 & 69.2 & 83.2 & 61.7 & 85.5 & 66.6 & 76.9 & 66.5 & 81.5 \\
            + {FP} & 66.8 & \best{64.6} & 68.9 & 72.0 & 61.3 & \best{51.6} & 64.4 & \best{65.0} & 65.3 & \best{63.3} & 67.6 & 69.6 & 67.7 & 67.2 & 60.1 & \best{70.0} & 63.9 & \best{62.7} & 64.8 & 67.4 \\
            \rowcolor{Gray}
            + {\sc DiffAdapt} & 69.2 & 68.0 & 71.4 & \best{67.2} & 63.3 & 60.6 & 68.6 & 82.5 & 68.1 & 69.6 & 69.1 & \best{65.1} & 71.1 & \best{38.0} & 63.4 & 79.8 & 68.1 & 83.3 & 67.9 & \best{66.6}  \\
            \midrule
            \multicolumn{21}{c}{\textbf{Blended trigger $\upuparrows$} \ \ \ \ \ \ \ \ \ \textbf{Perturbation trigger $\downdownarrows$}}\\
            \midrule
            Poisoning & 71.7 & 80.2 & 74.2 & 70.7 & 66.4 & 78.2 & 70.4 & 74.0 & 70.7 & 75.8 & 71.4 & 51.9 & 72.2 & 36.0 & 64.8 & 51.8 & 69.3 & 54.7 & 69.4 & 48.6 \\
            + {CLP} & 69.0 & 46.6 & 70.9 & 36.6 & 62.5 & 70.5 & 68.4 & 54.3 & 67.7 & 52.0 & 68.9 & 23.5 & 69.4 & 8.9  & 61.6 & 38.5 & 67.1 & 33.9 & 66.7 & 26.2 \\
            + {FP} & 67.7 & 43.8 & 69.3 & 32.1 & 61.5 & 43.4 & 65.2 & 33.5 & 65.9 & 38.2 & 67.4 & 19.3 & 68.1 & 6.5  & 60.8 & 24.0 & 64.2 & 15.9 & 65.1 & 16.4 \\
            \rowcolor{Gray}
            + {\sc DiffAdapt} & 68.9 & \best{1.5} & 71.3 & \best{0.9} & 63.4 & \best{4.8} & 67.5 & \best{3.5} & 67.8 & \best{2.6} & 69.4 & \best{4.5} & 70.9 & \best{0.1} & 64.1 & \best{2.0} & 68.0 & \best{0.2} & 68.1 & \best{1.7} \\
            \bottomrule
            \end{tabular}
        }
        \caption{ACC (\%) and ASR (\%) of {\sc DiffAdapt} against backdoor attacks on \textbf{OfficeHome} \citep{venkateswara2017deep} dataset for model adaptation (ResNet-50).} 
    \label{tab:officehome} 
    \end{table*}

\section{Experiment}

\subsection{Setup}

\textbf{Datasets.}
We evaluate our framework on three commonly used model adaptation benchmarks from image classification tasks.
\textbf{Office} \citep{saenko2010adapting} is a classic model adaptation dataset containing 31 categories across three domains (\ie, Amazon (A), DSLR (D), and Webcam (W)).
Since the small size of the data in the DSLR domain makes it difficult to poison a certain category, we remove two tasks whose target domain is DSLR and retain the remaining four (\ie, A$\to$W, D$\to$A, D$\to$W, W$\to$A).
\textbf{OfficeHome} \citep{venkateswara2017deep} is a popular dataset whose images are collected from office and home environments.
It consists of 65 categories across four domains (\ie, Art (A), Clipart (C), Product (P), and Real World (R)).
\textbf{DomainNet} \citep{peng2019moment} is a large-size challenging benchmark with imbalanced classes and extremely difficult tasks.
Following previous work \citep{tan2020class, li2021imbalanced}, we consider a subset version, \textbf{miniDomainNet} for convenience and efficiency.
miniDomainNet contains four domains (\ie, Clipart (C), Painting (P), Real (R), and Sketch (S)) and 40 categories.
For OfficeHome and miniDomainNet datasets, we use all 12 tasks to evaluate our framework.

\textbf{Evaluation metrics.}
In our experiments, we divide 80\% of the target domain samples as the unlabeled training set for adaptation and the remaining 20\% as the test set for metric calculation.
In order to avoid loss of generality, we uniformly select class 0 in alphabetical order as the target class for backdoor attack and defense.
We adopt accuracy on the clean samples (\textbf{ACC}) and attack success rate on the poison samples (\textbf{ASR}) to evaluate the effectiveness of our attack and defense method.
A stealthy attack should achieve high ASR while maintaining accuracy on clean samples to keep the backdoor from being detected.
Similarly, a great defense method should achieve both low ASR and high accuracy.

\textbf{Baselines.}
Since there are no methods designed for defending backdoor injection during the adaptation stage, we select two pruning techniques that can be combined with the model adaptation algorithms and require no annotated clean samples.
CLP \citep{zheng2022data} performs data-free pruning on the potentially backdoored model based on the upper bound of the channel Lipschitz constant.
FP \citep{liu2018fine} weakens backdoors by pruning the neurons with high average activation values.

\setlength{\tabcolsep}{4.0pt}
    \begin{table*}[!t]
        \centering
        \resizebox{0.95\textwidth}{!}{
            \begin{tabular}{l|rr|rr|rr|rr|rr|rr|rr|rr|rr|rr}
            \toprule
            & \multicolumn{10}{c|}{\textbf{SHOT} \citep{liang2020we}} & \multicolumn{10}{c}{\textbf{NRC} \citep{yang2021exploiting}} \\
            \midrule
            \multirow{2}{3em}{Task} & \multicolumn{2}{c|}{C$\to$} & \multicolumn{2}{c|}{P$\to$} & \multicolumn{2}{c|}{R$\to$} & \multicolumn{2}{c|}{S$\to$} & \multicolumn{2}{c|}{Avg} & \multicolumn{2}{c|}{C$\to$} & \multicolumn{2}{c|}{P$\to$} & \multicolumn{2}{c|}{R$\to$} & \multicolumn{2}{c|}{S$\to$} & \multicolumn{2}{c}{Avg}  \\ 
            & ACC & ASR & ACC & ASR & ACC & ASR & ACC & ASR & ACC & ASR & ACC & ASR & ACC & ASR & ACC & ASR & ACC & ASR & ACC & ASR  \\
            \midrule
            Source Only & 64.0 & \multicolumn{1}{c|}{-} & 70.2 & \multicolumn{1}{c|}{-} & 67.4 & \multicolumn{1}{c|}{-} & 66.8 & \multicolumn{1}{c|}{-} & 67.1 & \multicolumn{1}{c|}{-} & 64.0 & \multicolumn{1}{c|}{-}  & 70.2 & \multicolumn{1}{c|}{-} & 67.4 & \multicolumn{1}{c|}{-}  & 66.8 & \multicolumn{1}{c|}{-} & 67.1 & \multicolumn{1}{c}{-}  \\
            No Poisoning & 80.5 &  \multicolumn{1}{c|}{-} & 80.3 & \multicolumn{1}{c|}{-} & 77.5 & \multicolumn{1}{c|}{-} & 80.9 & \multicolumn{1}{c|}{-}  & 79.8 & \multicolumn{1}{c|}{-} & 80.7 & \multicolumn{1}{c|}{-}  & 81.6 & \multicolumn{1}{c|}{-}  & 77.5 & \multicolumn{1}{c|}{-}  & 83.2 & \multicolumn{1}{c|}{-} & 80.7 & \multicolumn{1}{c}{-} \\
            \midrule
            \midrule
            Poisoning & 78.8 & 63.3 & 79.6 & 44.5 & 76.3 & 26.0 & 80.6 & 46.9 & 78.8 & 45.2 & 80.6 & 59.2 & 80.8 & 43.1 & 77.1 & 20.9 & 83.1 & 78.1 & 80.4 & 50.3 \\
            + {CLP} & 76.2 & 65.2 & 77.9 & 44.4 & 73.9 & 26.2 & 78.7 & 56.3 & 76.7 & 48.0 & 78.4 & 62.7 & 78.7 & 47.4 & 74.1 & 28.2 & 80.7 & 82.4 & 78.0 & 55.2 \\
            + {FP} & 69.7 & 33.7 & 75.5 & 32.1 & 67.7 & 13.9 & 74.1 & 36.0 & 71.8 & 28.9 & 73.6 & \best{32.6} & 76.7 & \best{29.9} & 69.6 & 8.6  & 76.2 & 61.1 & 74.0 & 33.0 \\
            \rowcolor{Gray}
            + {\sc DiffAdapt} & 75.5 & \best{6.9} & 76.5 & \best{31.7} & 72.3 & \best{13.0} & 77.0 & \best{25.7} & 75.3 & \best{19.3} & 78.9 & 35.3 & 80.4 & 44.9 & 73.9 & \best{2.6} & 79.0 & \best{45.2} & 78.1 & \best{32.0} \\
            \midrule
            \multicolumn{21}{c}{\textbf{Blended trigger $\upuparrows$} \ \ \ \ \ \ \ \ \ \textbf{Perturbation trigger $\downdownarrows$}}\\
            \midrule
            Poisoning & 78.3 & 80.0 & 79.7 & 73.0 & 75.6 & 87.8 & 80.3 & 69.1 & 78.5 & 77.5 & 80.2 & 55.0 & 80.7 & 44.1 & 76.6 & 62.0 & 83.2 & 44.4 & 80.2 & 51.4 \\
            + {CLP} & 75.9 & 60.5 & 78.2 & 52.5 & 73.7 & 63.4 & 78.3 & 61.7 & 76.5 & 59.5 & 77.7 & 49.5 & 78.6 & 29.8 & 74.0 & 37.9 & 80.9 & 46.7 & 77.8 & 41.0 \\
            + {FP} & 69.3 & 31.4 & 75.8 & \best{14.4} & 67.9 & 34.2 & 73.5 & \best{29.7} & 71.6 & 27.4 & 73.7 & \best{9.1}  & 76.7 & \best{3.7}  & 70.3 & \best{6.5}  & 76.4 & 17.1 & 74.3 & \best{9.1}  \\
            \rowcolor{Gray}
            + {\sc DiffAdapt} & 74.0 & \best{8.0} & 76.7 & 26.1 & 72.7 & \best{14.3} & 77.0 & 32.5 & 75.1 & \best{20.2} & 78.0 & 12.1 & 80.1 & 15.3 & 74.7 & 14.2 & 78.7 & \best{16.4} & 77.9 & 14.5 \\
            \bottomrule
            \end{tabular}
        }
        \caption{ACC (\%) and ASR (\%) of {\sc DiffAdapt} against backdoor attacks on \textbf{miniDomainNet} \citep{peng2019moment} dataset for model adaptation (ResNet-50).}
    \label{tab:domainnet} 
    \end{table*}

\begin{figure*}[!t]
		\small
		\setlength\tabcolsep{1mm}
		\renewcommand\arraystretch{0.1}
		\begin{tabular}{cccc}
            \includegraphics[width=0.235\linewidth]{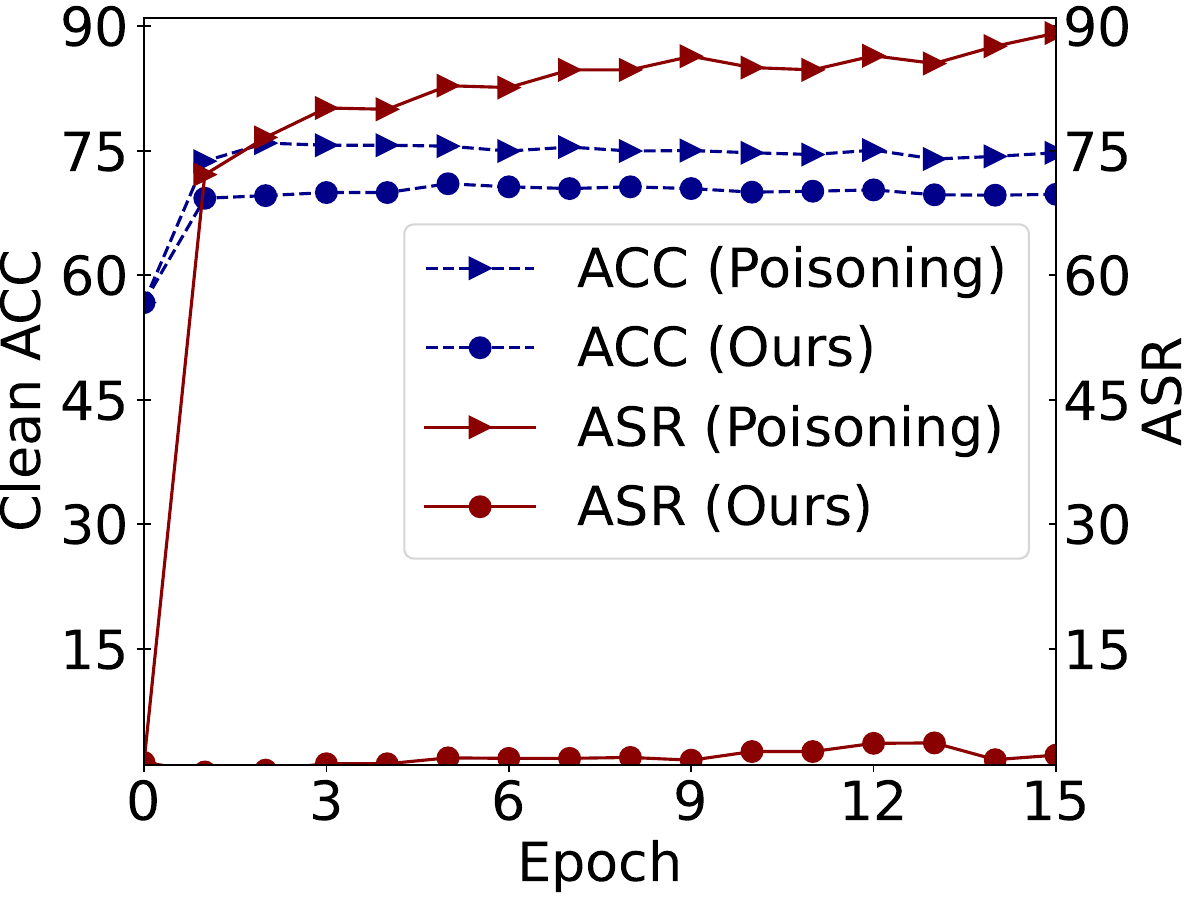} &		
            \includegraphics[width=0.235\linewidth, clip]{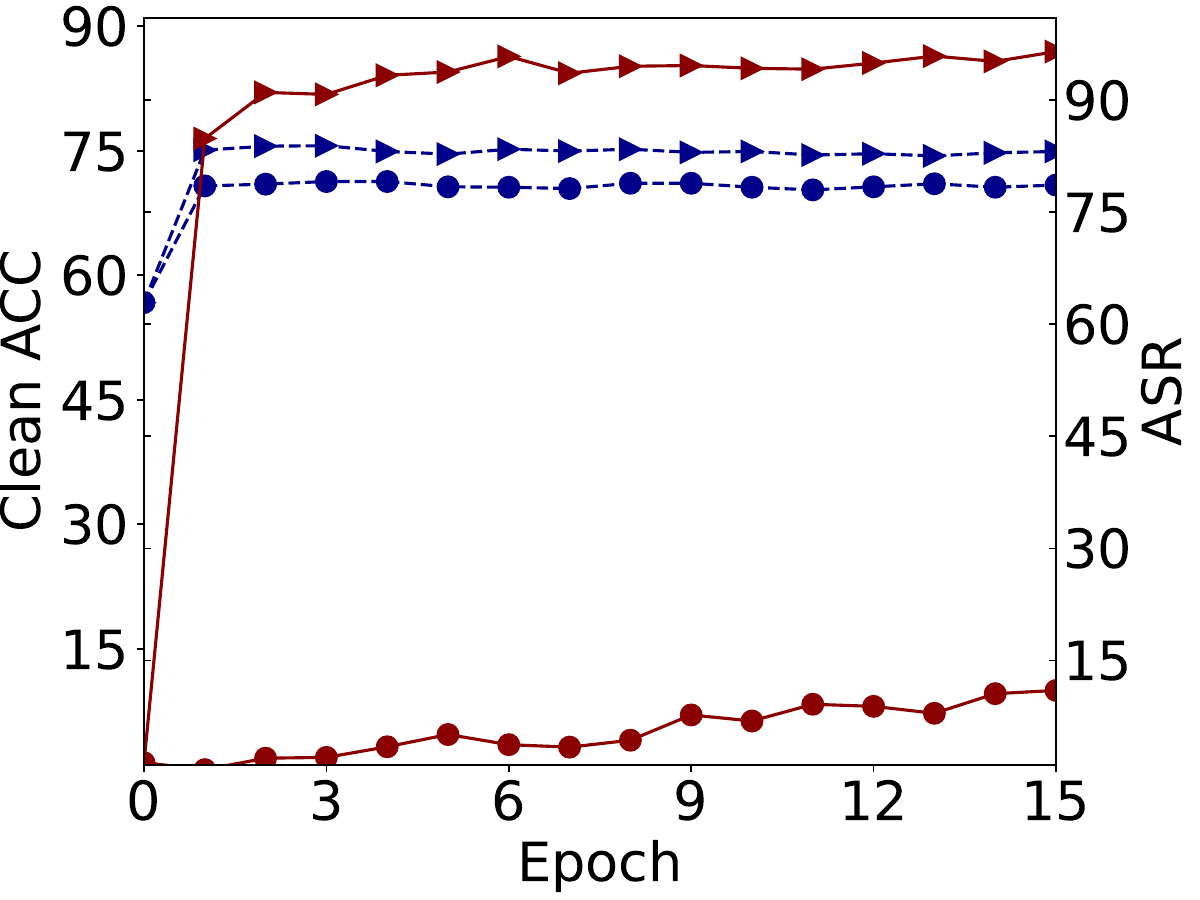} &
            \includegraphics[width=0.235\linewidth, clip]{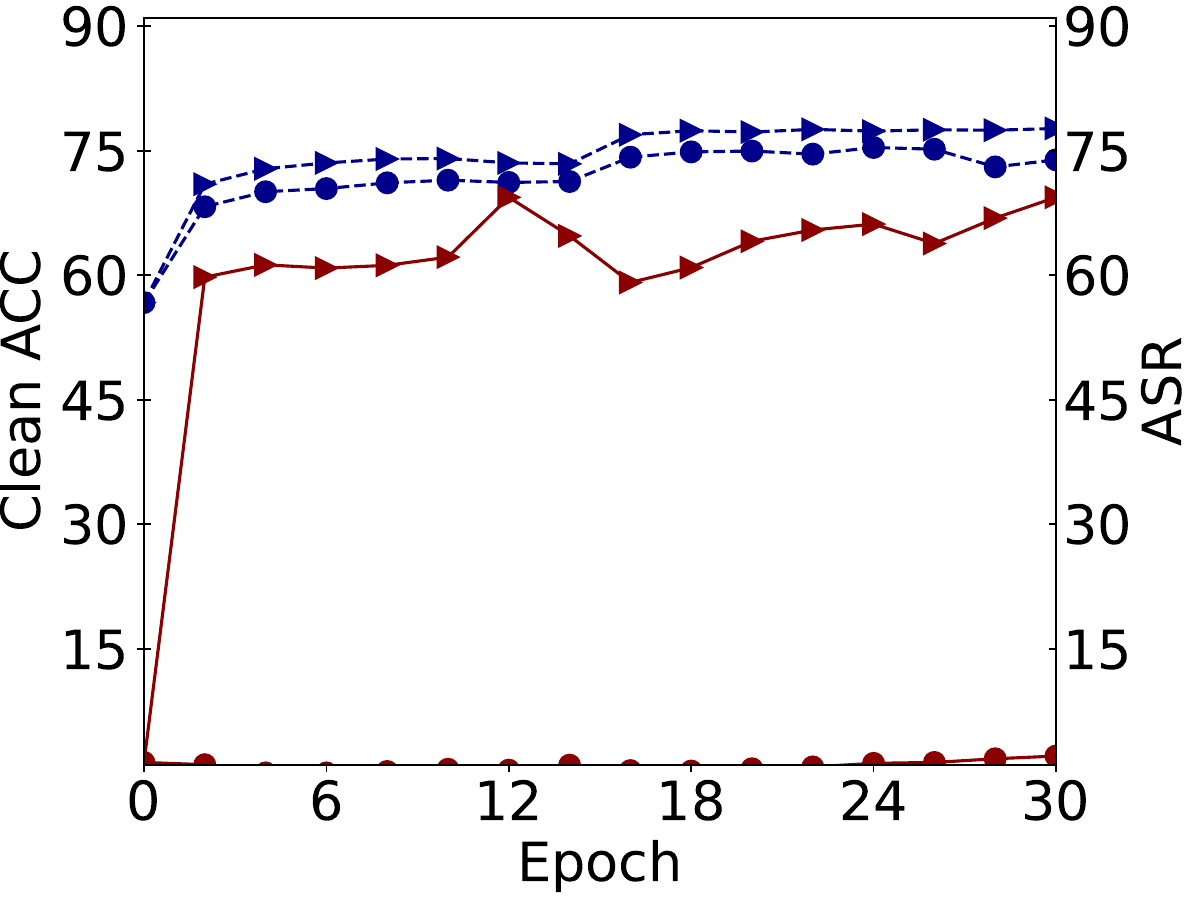} & 
            \includegraphics[width=0.235\linewidth, clip]{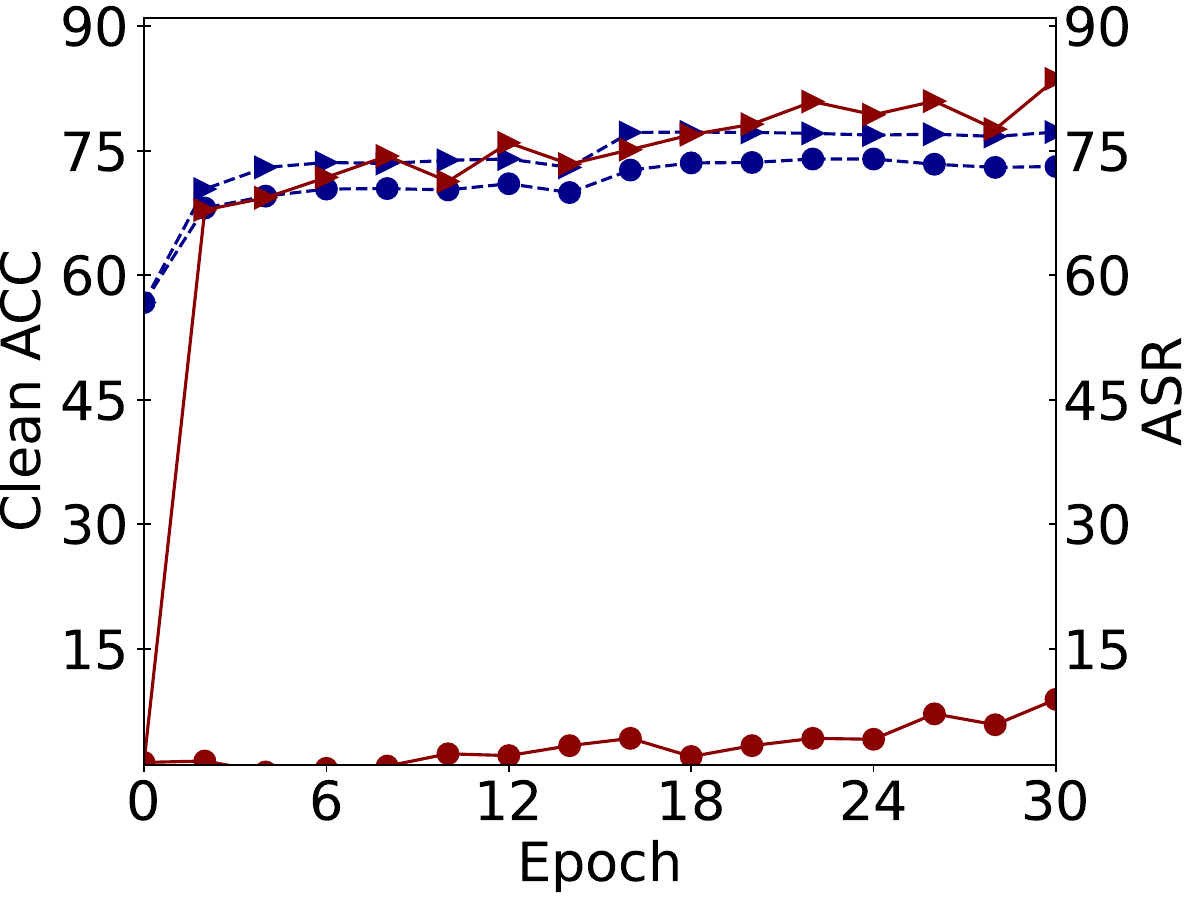} 
			\\[0.5mm]
			(a) SHOT (GT Strategy). & (b) SHOT (PL Strategy).  & (c) NRC (GT Strategy). & (d) NRC (PL Strategy).
		\end{tabular}
		\caption{ACC and ASR curve of backdoor attack and {\sc DiffAdapt} on C$\to$P from \textbf{miniDomainNet} \citep{peng2019moment}. }
		\label{fig:process}
\end{figure*}

\textbf{Implementation details.}
Different from supervised backdoor attacks, we choose two popular model adaptation methods, SHOT \citep{liang2020we} and NRC \citep{yang2021exploiting}, as victim algorithms.
We use their official codes and hyperparameters with ResNet-50 \citep{he2016deep}.
For each adaptation algorithm, we report the results from four attack methods (two types of trigger with two poison selection strategies).
For the non-optimization-based trigger, we use the Hello Kitty trigger in Blended \citep{chen2017targeted} directly.
The optimization-based trigger is implemented by GAP \citep{poursaeed2018generative} and $L_{inf}$ norm is 0.5 in a 120$\times$120 patch for Office and 100$\times$100 patch for others.

\textbf{Hyperparameters.}
For all experiments, we simply set the noise pixels to be sampled from a uniform distribution [-0.25, 0.25].
{\sc DiffAdapt} is a plug-and-play defense method for existing model adaptation algorithms, so no additional hyperparameters are introduced.
Other details are consistent with the official settings of the adaptation algorithms.

\subsection{Main Results}
We evaluate two backdoor triggers with the ground truth poisoning strategy and our defense method against the above attacks.
The results are shown in Table~\ref{tab:office}, \ref{tab:officehome}, \ref{tab:domainnet}.
Note that the results of the non-optimization-based (Blended) trigger are reported in the upper part of the tables and the results of the optimization-based backdoor (perturbation) trigger are provided in the lower part.
Due to space limitations, for OfficeHome and miniDomainNet, we report the average result across tasks from the same source domain and leave the detailed results in the \textbf{supplementary material}\footnote{Supplementary material is available at \url{https://github.com/TomSheng21/DiffAdapt/pdf/supp.pdf}.}.

\textbf{Results about backdoor attacks on model adaptation.}
Non-optimization-based backdoor attacks (Blended trigger) obtain a high ASR across various benchmarks.
Take results on OfficeHome in Table~\ref{tab:officehome} as an example, Blended trigger achieves an average ASR of 80.8\% on SHOT and 86.9\% on NRC.
Besides, the injection of Blended trigger maintains the target domain performance of the victim model which demonstrates its great concealment.
As shown in Table~\ref{tab:office}, \ref{tab:officehome}, compared with the clean training set, the poisoning set with Blended trigger only causes 0.6\% and 0.2\% decrease in accuracy on Office and OfficeHome datasets, respectively.

For optimization-based backdoor attacks, as shown in Table~\ref{tab:office}, the perturbation trigger achieves an average ASR of 81.6\% and 57.6\% on Office dataset on two adaptation algorithms.
And on miniDomainNet dataset in Table~\ref{tab:domainnet}, the average ASR of perturbation trigger on SHOT arrives at 77.5\%.
Also, the perturbation trigger keeps the model's performance, only reduces the clean accuracy of SHOT on miniDomainNet from 79.8\% to 78.5\% and causes a smaller 0.5\% gap on NRC algorithm.
Generally, the results demonstrate the effectiveness of two backdoor triggers, revealing the poisoning risk of unlabeled data during the adaptation.

\begin{figure}[t]
    \centering
    \includegraphics[width=0.95\linewidth]{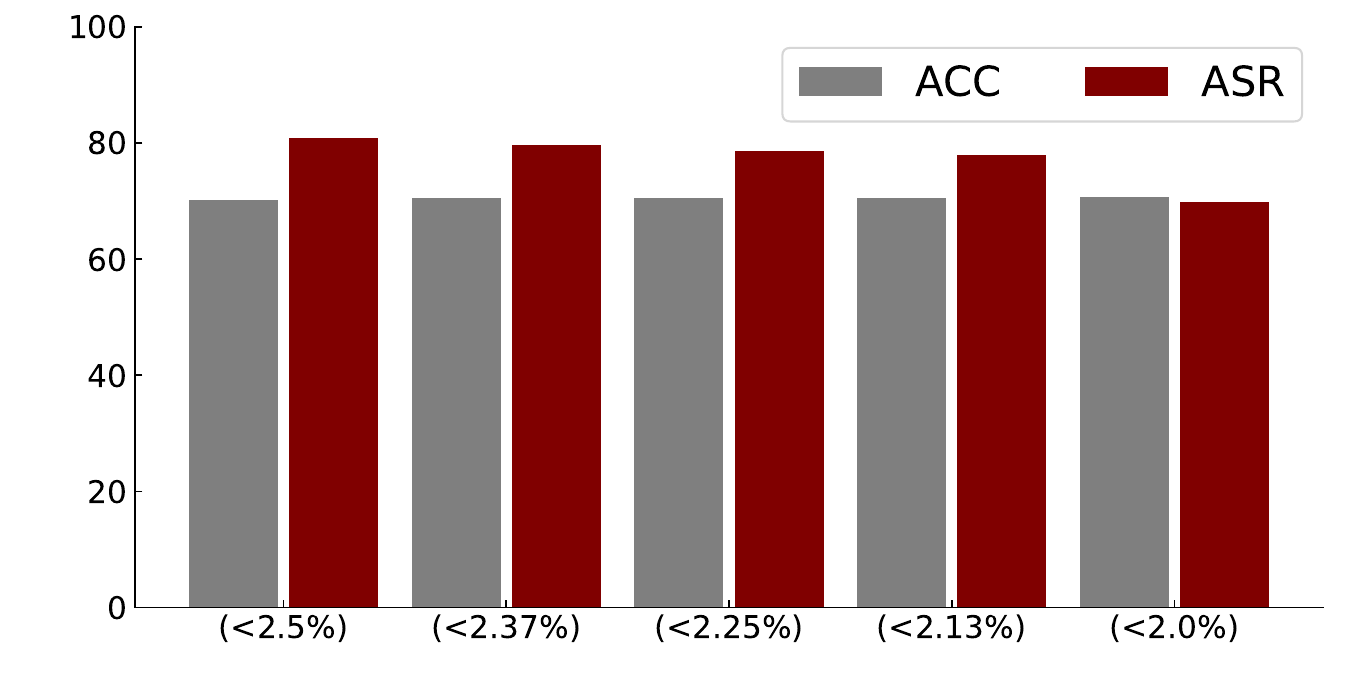}
    \caption{ACC (\%) and ASR (\%) of backdoor attacks under different poisoning rates for model adaptation.}
  \label{fig:rates}
\end{figure}

\setlength{\tabcolsep}{3.0pt}
    \begin{table}[!t]
        \centering
        \resizebox{0.45\textwidth}{!}{
            \begin{tabular}{l|cc|cc|cc|cc|cc}
            \toprule
            \multirow{2}{3em}{Task} &  \multicolumn{2}{c|}{A$\to$} & \multicolumn{2}{c|}{C$\to$} & \multicolumn{2}{c|}{P$\to$} & \multicolumn{2}{c|}{R$\to$} & \multicolumn{2}{c}{Avg}  \\ 
            & ACC & ASR & ACC & ASR & ACC & ASR & ACC & ASR & ACC & ASR \\
            \midrule
            Poisoning & 70.6 & 92.0 & 73.1 & 86.3 & 66.8 & 95.4 & 70.9 & 89.1 & 70.4 & 90.7 \\
            + {CLP} & 68.8 & 66.7 & 71.0 & 58.5 & 63.5 & 87.3 & 67.9 & 72.5 & 67.8 & 71.2 \\
            + {FP} & 66.9 & 60.3 & 68.3 & 43.7 & 62.7 & 80.4 & 65.7 & 51.0 & 65.9 & 58.9 \\
            \rowcolor{Gray}
            + {\sc DiffAdapt} & 68.4 & \best{1.4}  & 71.4 & \best{9.2}  & 64.4 & \best{7.4}  & 68.8 & \best{10.3} & 68.3 & \best{7.1} \\
            \bottomrule
            \end{tabular}
        }
        \caption{ACC (\%) and ASR (\%) of {\sc DiffAdapt} against backdoor attacks (\textbf{pseudo label strategy}) on \textbf{OfficeHome} \citep{venkateswara2017deep} dataset for model adaptation.}
    \label{tab:pl} 
    \end{table}

\textbf{Results about {\sc DiffAdapt} against backdoor attack.}
To defend against the backdoor attacks for the model adaptation, we further evaluate our proposed defense method {\sc DiffAdapt} on the above benchmarks, and the results are shown in Tables~\ref{tab:office}, \ref{tab:officehome}, \ref{tab:domainnet}.
It is obvious that {\sc DiffAdapt} effectively reduces ASR scores while maintaining the original classification ability.
Take Office dataset with perturbation trigger in Table~\ref{tab:office} as an example, {\sc DiffAdapt} reduces ASR scores from 81.6\% to 15.4\% on SHOT and from 57.6\% to 6.0\% on NRC.
At the same time, the clean accuracy of the target domain drops by 2.3\% and 2.2\% respectively, which is within an acceptable range.
Compared with baseline defense methods, {\sc DiffAdapt} always achieves better ASR and clean accuracy.
Although {\sc DiffAdapt} shows its superiority across most tasks, on OfficeHome data with the challenging Blended trigger, it is still slightly worse than FP whose defense is also weak.
In addition, we record the curves of ASR score and accuracy of backdoor attack and {\sc DiffAdapt} on the C$\to$P task from miniDomainNet in Fig.~\ref{fig:process}.
As expected, as shown in curves of adaptation, {\sc DiffAdapt} effectively defends against backdoor injection without affecting the convergence of the base algorithm.

\subsection{More Analysis}

\textbf{Analysis about pseudo labeling poisoning strategy.}
Besides the poisoning selection strategy based on ground truth labels, we also provide a pseudo-label poisoning strategy when the attacker can not access labels.
The results on OfficeHome dataset with the perturbation trigger are shown in Table~\ref{tab:pl}.
Pseudo-label-based strategy achieves a high ASR score of 90.7\% on SHOT algorithm on OfficeHome dataset.
Also, {\sc DiffAdapt} outperforms baseline methods, it reduces ASR from 90.7\% to 7.1\% and causes only a 2.1\% accuracy gap.
Although the CLP obtains slightly better accuracy, it produces poor performance at backdoor removal.
These results further demonstrate the flexibility of the proposed attack method and the effectiveness of {\sc DiffAdapt}.

\textbf{Analysis about different network architectures.}
To assess the versatility of our attack framework, we evaluate our attack method on a variety of network architectures including VGG, ViT, ConvNext, MobileNet, and ResNet101.
Since {\sc DiffAdapt} needs no well-designed modification for any networks, we also employ it on those backbones.
The results of Blended trigger on SHOT algorithm are provided in Table~\ref{tab:backbone}.
Take ConvNext for an example, our attack achieves an average ASR score of 72.5\%  while {\sc DiffAdapt} brings it down to 39.9\%.
It is shown that our attack method achieves a successful attack across different backbones and {\sc DiffAdapt} mitigates backdoor injection.

\textbf{Analysis about different poisoning rates.}
In most experiments, we assume that the attacker as well as the target domain provider can control the whole unlabeled dataset.
Here we study our attack method under various poison rates and report the results on OfficeHome with Blended trigger in Fig.~\ref{fig:rates}.
Please note that the rate in the figure refers to the poisoning rate of the whole target dataset and the samples of the target class are approximately equal to 2.5\%.
It is shown that a higher poisoning rate will result in a higher ASR score.
Moreover, various poisoning rates share almost the same performance on the clean sample.

\setlength{\tabcolsep}{4.0pt}
    \begin{table}[!t]
        \centering
        \resizebox{0.47\textwidth}{!}{
            \begin{tabular}{c|cc|cc|cc|cc|cc}
            \toprule
            {\multirow{2}{3em}{Task}} &  \multicolumn{2}{c|}{A$\to$} & \multicolumn{2}{c|}{C$\to$} & \multicolumn{2}{c|}{P$\to$} & \multicolumn{2}{c|}{R$\to$} & \multicolumn{2}{c}{Avg}  \\ 
            & ACC & ASR & ACC & ASR & ACC & ASR & ACC & ASR & ACC & ASR \\
            \midrule
            {VGG16} & 64.4 & 76.8 & 67.2 & 52.1 & 55.9 & 26.1 & 62.3 & 69.7 & 62.5 & 56.2 \\
             + {\sc DiffAdapt} & 63.0 & 24.5 & 65.1 & 6.6  & 52.9 & 3.7  & 58.6 & 38.3 & 59.9 & 18.3 \\
            \midrule
            {ViT-Base} & 78.3 & 50.7 & 56.1 & 32.5 & 73.2 & 65.5 & 75.4 & 44.7 & 70.7 & 48.4 \\
             + {\sc DiffAdapt} & 75.4 & 20.2 & 81.0 & 42.2 & 70.2 & 25.0 & 72.0 & 39.6 & 74.7 & 31.8 \\
            \midrule
            {ConvNext} & 80.0 & 65.7 & 84.8 & 78.5 & 74.9 & 77.5 & 77.6 & 68.2 & 79.3 & 72.5 \\
             + {\sc DiffAdapt} & 79.7 & 47.5 & 81.4 & 18.7 & 73.2 & 44.9 & 74.8 & 48.6 & 77.3 & 39.9 \\
            \midrule
            {MobileNet} & 63.6 & 80.7 & 62.8 & 56.1 & 57.2 & 26.8 & 63.7 & 77.1 & 61.8 & 60.2 \\
             + {\sc DiffAdapt} &  62.3 & 48.0 & 61.3 & 21.0 & 53.5 & 30.8 & 59.1 & 43.0 & 59.0 & 35.7 \\
            \midrule
            {Resnet101} & 73.2 & 86.4 & 76.2 & 88.7 & 68.6 & 88.8 & 73.1 & 87.8 & 72.8 & 87.9 \\
             + {\sc DiffAdapt} & 70.6 & 24.7 & 71.8 & 41.9 & 65.8 & 34.4 & 69.1 & 50.5 & 69.3 & 37.9 \\
            \bottomrule
            \end{tabular}
        }
        \caption{ACC (\%) and ASR (\%) with different backbones against backdoor attacks on \textbf{OfficeHome} dataset.}
    \label{tab:backbone} 
    \end{table}

\section{Conclusion}
This paper discusses whether users can trust unlabeled data during model adaptation.
Our study focuses on backdoor attacks during model adaptation and finds that a malicious data provider can achieve backdoor embedding through unsupervised poisoning.
Furthermore, to reduce the risks of potential backdoor attacks, we propose {\sc DiffAdapt}, a plug-and-play defense method to protect adaptation algorithms.
{\sc DiffAdapt} eliminates the association between triggers and target class by exchanging background areas among target samples.
Extensive experiments conducted on commonly used adaptation benchmarks validate the efficacy of {\sc DiffAdapt} in effectively defending against backdoor attacks.

\textbf{Limitation.}
It is worth noting that while our framework achieves successful attacks and defenses, some limitations still exist.
We only explore the classification problem, which is a relatively basic task.
Popular online adaptation algorithms employ minor optimization, making backdoor injection more difficult.
As for defense, compared to direct adaptation, {\sc DiffAdapt} requires twice the computational cost.
Improvements in these aspects can further improve the versatility of our framework.

\section*{Ethical Statement}
In our work, we explore trojan attack and defense in model adaptation, a sub-field of machine learning.
Given that model adaptation techniques are increasingly deployed in sensitive areas such as security systems and medical diagnostics, our proposed backdoor attack could potentially result in risky consequences, including accidents and security breaches.
To address these risks, we propose a robust defense framework aimed at helping adaptation algorithms defend against trojan attacks.
Our work highlights the vulnerability of model adaptation to such attacks, intending to raise awareness about these risks, particularly in high-stakes applications where the impact could be significant.

\section*{Acknowledgements}
This work was funded by the National Natural Science Foundation of China under Grants (62276256, U2441251) and the Young Elite Scientists Sponsorship Program by CAST (2023QNRC001).

\bibliography{main}

\begin{thebibliography}{56}
\providecommand{\natexlab}[1]{#1}

\bibitem[{Agarwal et~al.(2022)Agarwal, Paudel, Zaech, and Van~Gool}]{agarwal2022unsupervised}
Agarwal, P.; Paudel, D.~P.; Zaech, J.-N.; and Van~Gool, L. 2022.
\newblock Unsupervised robust domain adaptation without source data.
\newblock In \emph{Proc. WACV}, 2009--2018.

\bibitem[{Ahmed et~al.(2023)Ahmed, Al~Arafat, Rizve, Hossain, Guo, and Rakin}]{ahmed2023ssda}
Ahmed, S.; Al~Arafat, A.; Rizve, M.~N.; Hossain, R.; Guo, Z.; and Rakin, A.~S. 2023.
\newblock SSDA: Secure Source-Free Domain Adaptation.
\newblock In \emph{Proc. ICCV}, 19180--19190.

\bibitem[{Ben-David et~al.(2010)Ben-David, Blitzer, Crammer, Kulesza, Pereira, and Vaughan}]{ben2010theory}
Ben-David, S.; Blitzer, J.; Crammer, K.; Kulesza, A.; Pereira, F.; and Vaughan, J.~W. 2010.
\newblock A theory of learning from different domains.
\newblock \emph{Machine Learning}, 79(1): 151--175.

\bibitem[{Chen et~al.(2017)Chen, Liu, Li, Lu, and Song}]{chen2017targeted}
Chen, X.; Liu, C.; Li, B.; Lu, K.; and Song, D. 2017.
\newblock Targeted backdoor attacks on deep learning systems using data poisoning.
\newblock \emph{arXiv preprint arXiv:1712.05526}.

\bibitem[{Chou, Chen, and Ho(2023)}]{chou2023backdoor}
Chou, S.-Y.; Chen, P.-Y.; and Ho, T.-Y. 2023.
\newblock How to backdoor diffusion models?
\newblock In \emph{Proc. CVPR}, 4015--4024.

\bibitem[{Chou, Chen, and Ho(2024)}]{chou2024villandiffusion}
Chou, S.-Y.; Chen, P.-Y.; and Ho, T.-Y. 2024.
\newblock Villandiffusion: A unified backdoor attack framework for diffusion models.
\newblock In \emph{Proc. NeurIPS}.

\bibitem[{Dhariwal and Nichol(2021)}]{dhariwal2021diffusion}
Dhariwal, P.; and Nichol, A. 2021.
\newblock Diffusion models beat gans on image synthesis.
\newblock In \emph{Proc. NeurIPS}, 8780--8794.

\bibitem[{Ding et~al.(2023)Ding, Sheng, Liang, Zheng, and He}]{ding2023proxymix}
Ding, Y.; Sheng, L.; Liang, J.; Zheng, A.; and He, R. 2023.
\newblock Proxymix: Proxy-based mixup training with label refinery for source-free domain adaptation.
\newblock \emph{Neural Networks}, 167: 92--103.

\bibitem[{Doan et~al.(2021)Doan, Lao, Zhao, and Li}]{doan2021lira}
Doan, K.; Lao, Y.; Zhao, W.; and Li, P. 2021.
\newblock Lira: Learnable, imperceptible and robust backdoor attacks.
\newblock In \emph{Proc. ICCV}, 11966--11976.

\bibitem[{Dosovitskiy et~al.(2021)Dosovitskiy, Beyer, Kolesnikov, Weissenborn, Zhai, Unterthiner, Dehghani, Minderer, Heigold, Gelly et~al.}]{dosovitskiy2020image}
Dosovitskiy, A.; Beyer, L.; Kolesnikov, A.; Weissenborn, D.; Zhai, X.; Unterthiner, T.; Dehghani, M.; Minderer, M.; Heigold, G.; Gelly, S.; et~al. 2021.
\newblock An image is worth 16x16 words: Transformers for image recognition at scale.
\newblock In \emph{Proc. ICLR}.

\bibitem[{Fleuret et~al.(2021)}]{fleuret2021uncertainty}
Fleuret, F.; et~al. 2021.
\newblock Uncertainty reduction for model adaptation in semantic segmentation.
\newblock In \emph{Proc. CVPR}, 9613--9623.

\bibitem[{Ganin et~al.(2016)Ganin, Ustinova, Ajakan, Germain, Larochelle, Laviolette, Marchand, and Lempitsky}]{ganin2016domain}
Ganin, Y.; Ustinova, E.; Ajakan, H.; Germain, P.; Larochelle, H.; Laviolette, F.; Marchand, M.; and Lempitsky, V. 2016.
\newblock Domain-adversarial training of neural networks.
\newblock \emph{Machine Learning}, 17(1): 2096--2030.

\bibitem[{Gu, Dolan-Gavitt, and Garg(2017)}]{gu2017badnets}
Gu, T.; Dolan-Gavitt, B.; and Garg, S. 2017.
\newblock Badnets: Identifying vulnerabilities in the machine learning model supply chain.
\newblock \emph{arXiv preprint arXiv:1708.06733}.

\bibitem[{Guan, Liang, and He(2024)}]{guan2024backdoor}
Guan, J.; Liang, J.; and He, R. 2024.
\newblock Backdoor Defense via Test-Time Detecting and Repairing.
\newblock In \emph{Proc. CVPR}, 24564--24573.

\bibitem[{Guan et~al.(2022)Guan, Tu, He, and Tao}]{guan2022few}
Guan, J.; Tu, Z.; He, R.; and Tao, D. 2022.
\newblock Few-shot backdoor defense using shapley estimation.
\newblock In \emph{Proc. CVPR}, 13358--13367.

\bibitem[{He et~al.(2016)He, Zhang, Ren, and Sun}]{he2016deep}
He, K.; Zhang, X.; Ren, S.; and Sun, J. 2016.
\newblock Deep residual learning for image recognition.
\newblock In \emph{Proc. CVPR}, 770--778.

\bibitem[{Huang et~al.(2021)Huang, Guan, Xiao, and Lu}]{huang2021model}
Huang, J.; Guan, D.; Xiao, A.; and Lu, S. 2021.
\newblock Model adaptation: Historical contrastive learning for unsupervised domain adaptation without source data.
\newblock In \emph{Proc. NeurIPS}, 3635--3649.

\bibitem[{Krizhevsky, Sutskever, and Hinton(2012)}]{krizhevsky2012imagenet}
Krizhevsky, A.; Sutskever, I.; and Hinton, G.~E. 2012.
\newblock Imagenet classification with deep convolutional neural networks.
\newblock In \emph{Proc. NeurIPS}.

\bibitem[{Li et~al.(2023)Li, Pang, Xi, Du, Ji, Yao, and Wang}]{li2023embarrassingly}
Li, C.; Pang, R.; Xi, Z.; Du, T.; Ji, S.; Yao, Y.; and Wang, T. 2023.
\newblock An Embarrassingly Simple Backdoor Attack on Self-supervised Learning.
\newblock In \emph{Proc. ICCV}, 4367--4378.

\bibitem[{Li et~al.(2020)Li, Jiao, Cao, Wong, and Wu}]{li2020model}
Li, R.; Jiao, Q.; Cao, W.; Wong, H.-S.; and Wu, S. 2020.
\newblock Model adaptation: Unsupervised domain adaptation without source data.
\newblock In \emph{Proc. CVPR}, 9641--9650.

\bibitem[{Li et~al.(2021{\natexlab{a}})Li, Chen, Xie, Yang, Yuan, Pu, and Zhuang}]{li2021free}
Li, X.; Chen, W.; Xie, D.; Yang, S.; Yuan, P.; Pu, S.; and Zhuang, Y. 2021{\natexlab{a}}.
\newblock A free lunch for unsupervised domain adaptive object detection without source data.
\newblock In \emph{Proc. AAAI}, 8474--8481.

\bibitem[{Li et~al.(2021{\natexlab{b}})Li, Li, Zhu, Wang, and Huang}]{li2021imbalanced}
Li, X.; Li, J.; Zhu, L.; Wang, G.; and Huang, Z. 2021{\natexlab{b}}.
\newblock Imbalanced source-free domain adaptation.
\newblock In \emph{Proc. ACM MM}, 3330--3339.

\bibitem[{Li et~al.(2022)Li, Jiang, Li, and Xia}]{li2022backdoor}
Li, Y.; Jiang, Y.; Li, Z.; and Xia, S.-T. 2022.
\newblock Backdoor learning: A survey.
\newblock \emph{IEEE Transactions on Neural Networks and Learning Systems}, 35(1): 5--22.

\bibitem[{Li et~al.(2021{\natexlab{c}})Li, Li, Wu, Li, He, and Lyu}]{li2021invisible}
Li, Y.; Li, Y.; Wu, B.; Li, L.; He, R.; and Lyu, S. 2021{\natexlab{c}}.
\newblock Invisible backdoor attack with sample-specific triggers.
\newblock In \emph{Proc. ICCV}, 16463--16472.

\bibitem[{Li et~al.(2021{\natexlab{d}})Li, Lyu, Koren, Lyu, Li, and Ma}]{li2021neural}
Li, Y.; Lyu, X.; Koren, N.; Lyu, L.; Li, B.; and Ma, X. 2021{\natexlab{d}}.
\newblock Neural attention distillation: Erasing backdoor triggers from deep neural networks.
\newblock In \emph{Proc. ICLR}.

\bibitem[{Liang, He, and Tan(2024)}]{liang2024comprehensive}
Liang, J.; He, R.; and Tan, T. 2024.
\newblock A comprehensive survey on test-time adaptation under distribution shifts.
\newblock \emph{International Journal of Computer Vision}, 1--34.

\bibitem[{Liang, Hu, and Feng(2020)}]{liang2020we}
Liang, J.; Hu, D.; and Feng, J. 2020.
\newblock Do we really need to access the source data? source hypothesis transfer for unsupervised domain adaptation.
\newblock In \emph{Proc. ICML}, 6028--6039.

\bibitem[{Liang et~al.(2021{\natexlab{a}})Liang, Hu, Feng, and He}]{liang2021umad}
Liang, J.; Hu, D.; Feng, J.; and He, R. 2021{\natexlab{a}}.
\newblock Umad: Universal model adaptation under domain and category shift.
\newblock \emph{arXiv preprint arXiv:2112.08553}.

\bibitem[{Liang et~al.(2022)Liang, Hu, Feng, and He}]{liang2022dine}
Liang, J.; Hu, D.; Feng, J.; and He, R. 2022.
\newblock Dine: Domain adaptation from single and multiple black-box predictors.
\newblock In \emph{Proc. CVPR}, 8003--8013.

\bibitem[{Liang et~al.(2021{\natexlab{b}})Liang, Hu, Wang, He, and Feng}]{liang2021source}
Liang, J.; Hu, D.; Wang, Y.; He, R.; and Feng, J. 2021{\natexlab{b}}.
\newblock Source data-absent unsupervised domain adaptation through hypothesis transfer and labeling transfer.
\newblock \emph{IEEE Trans on Pattern Analysis and Machine Intelligence}, 44(11): 8602--8617.

\bibitem[{Liao et~al.(2024)Liao, Yi, Shi, Yang, Fang, and Yang}]{liao2024imperceptible}
Liao, J.; Yi, L.; Shi, W.; Yang, W.; Fang, Y.; and Yang, X. 2024.
\newblock Imperceptible backdoor watermarks for speech recognition model copyright protection.
\newblock \emph{Visual Intelligence}, 2(1): 23.

\bibitem[{Liu, Dolan-Gavitt, and Garg(2018)}]{liu2018fine}
Liu, K.; Dolan-Gavitt, B.; and Garg, S. 2018.
\newblock Fine-pruning: Defending against backdooring attacks on deep neural networks.
\newblock In \emph{International Symposium on Research in Attacks, Intrusions and Defenses}, 273--294.

\bibitem[{Liu, Zhang, and Wang(2021)}]{liu2021source}
Liu, Y.; Zhang, W.; and Wang, J. 2021.
\newblock Source-free domain adaptation for semantic segmentation.
\newblock In \emph{Proc. CVPR}, 1215--1224.

\bibitem[{Long et~al.(2018)Long, Cao, Wang, and Jordan}]{long2018conditional}
Long, M.; Cao, Z.; Wang, J.; and Jordan, M.~I. 2018.
\newblock Conditional adversarial domain adaptation.
\newblock In \emph{Proc. NeurIPS}.

\bibitem[{Nguyen and Tran(2021)}]{nguyen2021wanet}
Nguyen, A.; and Tran, A. 2021.
\newblock WaNet--Imperceptible Warping-based Backdoor Attack.
\newblock In \emph{Proc. ICLR}.

\bibitem[{Peng et~al.(2019)Peng, Bai, Xia, Huang, Saenko, and Wang}]{peng2019moment}
Peng, X.; Bai, Q.; Xia, X.; Huang, Z.; Saenko, K.; and Wang, B. 2019.
\newblock Moment matching for multi-source domain adaptation.
\newblock In \emph{Proc. ICCV}, 1406--1415.

\bibitem[{Poursaeed et~al.(2018)Poursaeed, Katsman, Gao, and Belongie}]{poursaeed2018generative}
Poursaeed, O.; Katsman, I.; Gao, B.; and Belongie, S. 2018.
\newblock Generative adversarial perturbations.
\newblock In \emph{Proc. CVPR}.

\bibitem[{Radford et~al.(2021)Radford, Kim, Hallacy, Ramesh, Goh, Agarwal, Sastry, Askell, Mishkin, Clark et~al.}]{radford2021learning}
Radford, A.; Kim, J.~W.; Hallacy, C.; Ramesh, A.; Goh, G.; Agarwal, S.; Sastry, G.; Askell, A.; Mishkin, P.; Clark, J.; et~al. 2021.
\newblock Learning transferable visual models from natural language supervision.
\newblock In \emph{Proc. ICML}, 8748--8763.

\bibitem[{Saenko et~al.(2010)Saenko, Kulis, Fritz, and Darrell}]{saenko2010adapting}
Saenko, K.; Kulis, B.; Fritz, M.; and Darrell, T. 2010.
\newblock Adapting visual category models to new domains.
\newblock In \emph{Proc. ECCV}, 213--226.

\bibitem[{Saha et~al.(2022)Saha, Tejankar, Koohpayegani, and Pirsiavash}]{saha2022backdoor}
Saha, A.; Tejankar, A.; Koohpayegani, S.~A.; and Pirsiavash, H. 2022.
\newblock Backdoor attacks on self-supervised learning.
\newblock In \emph{Proc. CVPR}, 13337--13346.

\bibitem[{Shejwalkar, Lyu, and Houmansadr(2023)}]{shejwalkar2023perils}
Shejwalkar, V.; Lyu, L.; and Houmansadr, A. 2023.
\newblock The perils of learning from unlabeled data: Backdoor attacks on semi-supervised learning.
\newblock In \emph{Proc. ICCV}, 4730--4740.

\bibitem[{Sheng et~al.(2023)Sheng, Liang, He, Wang, and Tan}]{sheng2023adaptguard}
Sheng, L.; Liang, J.; He, R.; Wang, Z.; and Tan, T. 2023.
\newblock AdaptGuard: Defending Against Universal Attacks for Model Adaptation.
\newblock In \emph{Proc. ICCV}.

\bibitem[{Tan, Peng, and Saenko(2020)}]{tan2020class}
Tan, S.; Peng, X.; and Saenko, K. 2020.
\newblock Class-imbalanced domain adaptation: an empirical odyssey.
\newblock In \emph{Proc. ECCV Workshops}, 585--602.

\bibitem[{Tian et~al.(2021)Tian, Zhang, Li, and Xu}]{tian2021vdm}
Tian, J.; Zhang, J.; Li, W.; and Xu, D. 2021.
\newblock VDM-DA: Virtual domain modeling for source data-free domain adaptation.
\newblock \emph{IEEE Transactions on Circuits and Systems for Video Technology}, 32(6): 3749--3760.

\bibitem[{Venkateswara et~al.(2017)Venkateswara, Eusebio, Chakraborty, and Panchanathan}]{venkateswara2017deep}
Venkateswara, H.; Eusebio, J.; Chakraborty, S.; and Panchanathan, S. 2017.
\newblock Deep hashing network for unsupervised domain adaptation.
\newblock In \emph{Proc. CVPR}, 5018--5027.

\bibitem[{Wang et~al.(2019)Wang, Yao, Shan, Li, Viswanath, Zheng, and Zhao}]{wang2019neural}
Wang, B.; Yao, Y.; Shan, S.; Li, H.; Viswanath, B.; Zheng, H.; and Zhao, B.~Y. 2019.
\newblock Neural cleanse: Identifying and mitigating backdoor attacks in neural networks.
\newblock In \emph{Proc. S\&P}, 707--723.

\bibitem[{Wang et~al.(2021)Wang, Shelhamer, Liu, Olshausen, and Darrell}]{wang2020tent}
Wang, D.; Shelhamer, E.; Liu, S.; Olshausen, B.; and Darrell, T. 2021.
\newblock Tent: Fully test-time adaptation by entropy minimization.
\newblock In \emph{Proc. ICLR}.

\bibitem[{Wu et~al.(2022)Wu, Chen, Zhang, Zhu, Wei, Yuan, and Shen}]{wu2022backdoorbench}
Wu, B.; Chen, H.; Zhang, M.; Zhu, Z.; Wei, S.; Yuan, D.; and Shen, C. 2022.
\newblock Backdoorbench: A comprehensive benchmark of backdoor learning.
\newblock In \emph{Proc. NeurIPS}, 10546--10559.

\bibitem[{Wu and Wang(2021)}]{wu2021adversarial}
Wu, D.; and Wang, Y. 2021.
\newblock Adversarial neuron pruning purifies backdoored deep models.
\newblock In \emph{Proc. NeurIPS}, 16913--16925.

\bibitem[{Yang et~al.(2021)Yang, van~de Weijer, Herranz, Jui et~al.}]{yang2021exploiting}
Yang, S.; van~de Weijer, J.; Herranz, L.; Jui, S.; et~al. 2021.
\newblock Exploiting the intrinsic neighborhood structure for source-free domain adaptation.
\newblock In \emph{Proc. NeurIPS}, 29393--29405.

\bibitem[{Yu et~al.(2023)Yu, Sheng, He, and Liang}]{yu2023benchmarking}
Yu, Y.; Sheng, L.; He, R.; and Liang, J. 2023.
\newblock Benchmarking test-time adaptation against distribution shifts in image classification.
\newblock \emph{arXiv preprint arXiv:2307.03133}.

\bibitem[{Yu et~al.(2024)Yu, Sheng, He, and Liang}]{yu2024stamp}
Yu, Y.; Sheng, L.; He, R.; and Liang, J. 2024.
\newblock STAMP: Outlier-Aware Test-Time Adaptation with Stable Memory Replay.
\newblock In \emph{Proc. ECCV}, 375--392.

\bibitem[{Zhang et~al.(2023{\natexlab{a}})Zhang, Huang, Jiang, and Lu}]{zhang2023black}
Zhang, J.; Huang, J.; Jiang, X.; and Lu, S. 2023{\natexlab{a}}.
\newblock Black-box Unsupervised Domain Adaptation with Bi-directional Atkinson-Shiffrin Memory.
\newblock In \emph{Proc. ICCV}, 11771--11782.

\bibitem[{Zhang et~al.(2022)Zhang, Chen, Cheng, Li, Li, Lin, and Li}]{zhang2022divide}
Zhang, Z.; Chen, W.; Cheng, H.; Li, Z.; Li, S.; Lin, L.; and Li, G. 2022.
\newblock Divide and contrast: Source-free domain adaptation via adaptive contrastive learning.
\newblock In \emph{Proc. NeurIPS}, 5137--5149.

\bibitem[{Zhang et~al.(2023{\natexlab{b}})Zhang, Xiao, Li, Lv, Qi, Liu, Wang, Jiang, and Sun}]{zhang2023red}
Zhang, Z.; Xiao, G.; Li, Y.; Lv, T.; Qi, F.; Liu, Z.; Wang, Y.; Jiang, X.; and Sun, M. 2023{\natexlab{b}}.
\newblock Red alarm for pre-trained models: Universal vulnerability to neuron-level backdoor attacks.
\newblock \emph{Machine Intelligence Research}, 20(2): 180--193.

\bibitem[{Zheng et~al.(2022)Zheng, Tang, Li, and Liu}]{zheng2022data}
Zheng, R.; Tang, R.; Li, J.; and Liu, L. 2022.
\newblock Data-free backdoor removal based on channel lipschitzness.
\newblock In \emph{Proc. ECCV}, 175--191.

\end{thebibliography}

\end{document}